\documentclass[traditabstract,twocolumns]{aa} 

\usepackage{graphicx}
\usepackage{epstopdf}
\usepackage{deluxetable}
\usepackage{txfonts,epsfig} 
\usepackage{mathrsfs}  
\usepackage{natbib}
\usepackage{amssymb}
\bibpunct{(}{)}{;}{a}{}{,} 
\usepackage{longtable}
\usepackage{caption}
\usepackage{url}
\usepackage{hyperref}

\begin{document}

\authorrunning{Taddia, Moquist, Sollerman, et al.}
\titlerunning{}

\title{Metallicity from Type II Supernovae from the (i)PTF}

\author{F. Taddia\inst{1},
 P. Moquist\inst{1},
  J. Sollerman\inst{1},
 A. Rubin\inst{2},
 G. Leloudas\inst{2,3},
  A. Gal-Yam\inst{2},
  I. Arcavi\inst{4,5},
 Y. Cao\inst{6},
 A.~V. Filippenko\inst{7},
 M.~L. Graham\inst{7},
 P.~A. Mazzali\inst{8,9},
 P.~E. Nugent\inst{7,10},
 Y.-C. Pan\inst{11},
J.~M. Silverman\inst{12},
 D. Xu\inst{13},
  O. Yaron\inst{2}}
 
\institute{The Oskar Klein Centre, Department of Astronomy, Stockholm University, AlbaNova, 10691 Stockholm, Sweden. 
\and Department of Particle Physics \& Astrophysics, Weizmann Institute of Science, Rehovot 76100, Israel.
\and Dark Cosmology Centre, Niels Bohr Institute, University of Copenhagen, Juliane Maries Vej 30, 2100 Copenhagen, Denmark.
\and Las Cumbres Observatory Global Telescope, 6740 Cortona Dr, Suite 102, Goleta, CA 93117, USA.
\and Kavli Institute for Theoretical Physics, University of California, Santa Barbara, CA 93106, USA.
\and Astronomy Department, California Institute of Technology, Pasadena, California 91125, USA.
\and Department of Astronomy, University of California, Berkeley, CA 94720-3411, USA.
\and Astrophysics Research Institute, Liverpool John Moores University, 146 Brownlow Hill, Liverpool L3 5RF, UK.
\and Max-Planck-Institut f\"{u}r Astrophysik, Karl-Schwarzschild-Str. 1, D-85748 Garching, Germany.
\and Lawrence Berkeley National Laboratory, Berkeley, CA, 94720, USA.
\and Astronomy Department, University of Illinois at Urbana-Champaign, 1002 W. Green Street, Urbana, IL 61801, USA.
\and Department of Astronomy, University of Texas, Austin, TX 78712, USA.\and Key Laboratory of Space Astronomy and Technology, National Astronomical Observatories, Chinese Academy of Sciences, 20A Datun Road, Beijing 100012, China.}

\date{Received; accepted}

\abstract{
Type IIP supernovae (SNe~IIP) have recently been proposed as metallicity ($Z$) probes. The spectral models of Dessart et al. (2014) showed that the pseudo-equivalent width of \ion{Fe}{ii}~$\lambda$5018 (pEW$_{5018}$) during the plateau phase depends on the primordial $Z$, but there was a paucity of SNe~IIP exhibiting pEW$_{5018}$ that were compatible with $Z~<~0.4~{\rm Z}_{\odot}$. This lack might be due to some physical property of the SN~II population or to the fact that those SNe have been discovered in luminous, metal-rich targeted galaxies. 
Here we use SN~II observations from the untargeted  (intermediate) Palomar Transient Factory [(i)PTF] survey, aiming to investigate the pEW$_{5018}$ distribution of this SN population and, in particular, to look for the presence of SNe~II at lower $Z$. 
We perform pEW$_{5018}$ measurements on the spectra of a sample of 39 (i)PTF SNe~II, selected to have well-constrained explosion epochs and light-curve properties. Based on the comparison with the pEW$_{5018}$ spectral models, we subgrouped our SNe into four $Z$ bins from $Z~\approx~0.1$~$Z_{\odot}$ up to $Z~\approx~2$~$Z_{\odot}$. We also independently investigated the $Z$ of the hosts by using their absolute magnitudes and colors and, in a few cases, using strong-line diagnostics from spectra. We searched for possible correlations between SN observables, such as their peak magnitudes and the $Z$ inferred from pEW$_{5018}$. 
We found 11 events with pEW$_{5018}$ that were small enough to indicate $Z~\approx~0.1$~$Z_{\odot}$. The trend of pEW$_{5018}$ with $Z$ matches the $Z$ estimates obtained from the host-galaxy photometry, although the significance of the correlation is weak. We also found that SNe with brighter peak magnitudes have smaller pEW$_{5018}$ and occur at lower $Z$.}



\maketitle

\section{Introduction}
\label{sec:intro}
\begin{figure*}
\includegraphics[width=18cm]{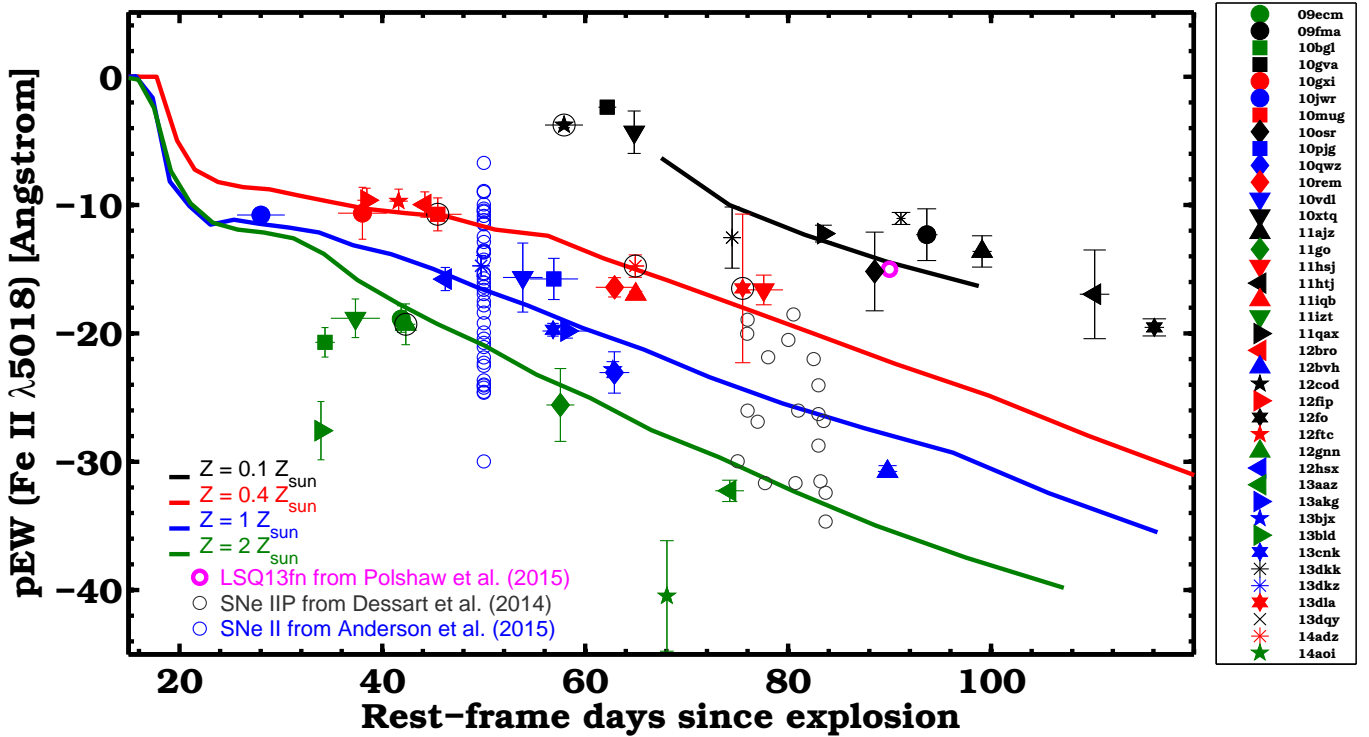}
\caption{\label{EW}pEW$_{5018}$ as a function of the SN phase. Filled symbols label our (i)PTF~SNe~II. 
Most of them are SNe~IIP, 
and those circled in black are SNe~IIL (i.e., they show a decline rate $>~0.014$~mag~d$^{-1}$ on the plateau). The spectral models by D14 indicating different $Z$ are represented by solid lines. Our SNe are subgrouped and color-coded in 4 subsets based on the pEW(t) model to which they appear closest.
pEW measurements from D14 and \citet{anderson15a} are shown by empty circles (gray and blue, respectively). Our SN sample includes a subset of objects with unprecedentedly small pEW, consistent with $Z_{\rm SN}~=~0.1$~$Z_{\odot}$. The only object with similar pEW(\ion{Fe}{ii}~5018) is LSQ13fn \citep{polshaw15}.}
\end{figure*}
Type II supernovae (SNe) are the most common core-collapse SN  events \citep{li11}. They are characterized by hydrogen-rich spectra \citep[e.g.,][]{filippenko97}, and their light curves exhibit a fast rise to peak \citep[][hereafter R15]{rubin16}, followed by a long ($\sim90$~d) plateau in the case of SNe~IIP or by a linear decline ($>1.4$~mag/100~d) in the case of SNe~IIL. \citet{anderson14} show that these two subclasses 
may actually be the extremes of a continuum, with several objects showing intermediate light-curve slopes. The nature of the progenitors of SNe~IIP is well established: pre-explosion images at their locations show extended ($R~\gtrsim~500$~R$_{\odot}$) red supergiants (RSGs) in the mass range between 8.5 and 17~M$_{\odot}$ \citep{smartt09}.

Recently, \citet[][hereafter D14]{dessart14} have proposed the use of SNe~II as metallicity ($Z$) probes. In their work, SN~II spectral models (first presented in \citealp{dessart13}) show that the equivalent width (EW) of metal lines such as \ion{Fe}{ii}~$\lambda\lambda$5018, 5169 depends on the $Z$ of the SN progenitor, as well as on the spectral phase. Also, the pseudo-EW (pEW) of these lines, which is more easily measurable than the actual EW, is a function of $Z$ and phase. 
D14 measured the pEW of \ion{Fe}{ii}~$\lambda$5018 [hereafter pEW$_{5018}$] in SN~IIP spectra during the plateau phase and compared it with the pEW$_{5018}$ of their spectral models in order to determine the $Z$ at the SN locations. \ion{Fe}{ii}~$\lambda$5018 was chosen because it is easy to observe in SN~II spectra and is less affected by line blending than the stronger \ion{Fe}{ii}~$\lambda$5169 line, whose pEW is also a proxy for $Z$. 
\citet{anderson15a} recently presented ongoing investigations of the correlation between the pEW$_{5018}$ and the SN progenitor $Z$ as measured from the emission lines of 43 SN~II host galaxies, at least in the range between 12 $+$ log(O/H)~$=~8.2$--8.6.

Using spectral data mainly from the Carnegie Supernova Project (CSP), D14 suggest that there is a lack of SNe~IIP at $Z~\lesssim 0.4$~$Z_{\odot}$. This could 
be a characteristic of the SN~IIP population, thus providing clues to their progenitor evolution and explosion mechanisms. However, it could also
be a bias effect, since the CSP 
mainly observed SNe that were discovered by targeting luminous and therefore 
metal-rich galaxies. 
 \citet{anderson15a} also show a lack of SNe~II with small pEW$_{5018}$ --- that is, at low $Z$ (see their figure~1a).  LSQ13fn \citep{polshaw15} shows a small pEW$_{5018}$, corresponding to $Z~\approx~0.1$~$Z_{\odot}$, but it seems to reside in a solar-$Z$ host galaxy. (On the other hand, it has a large projected offset from the host-galaxy center.)
 Whether the lack of SNe~II at low $Z$ is a bias effect or a property of this SN class can only be tested with a larger sample of events discovered by an untargeted survey. 

The Palomar Transient Factory (PTF) and its continuation (the intermediate PTF) are untargeted surveys, which allowed the discovery of a large number of core-collapse SNe in a wide variety of galaxies. 
\citet{arcavi10} studied the PTF~SN populations in dwarf galaxies, finding an excess of SNe~IIb as compared to the SN population in brighter hosts.  
Thanks to the high cadence of PTF and iPTF [hereafter (i)PTF], for many targets there are also good constraints on the explosion epoch.
Furthermore, the (i)PTF collaboration has access to many telescopes for SN follow-up observations \citep{galyam11}, and it has collected a large number of high signal-to-noise (S/N) ratio SN spectra that are needed to study the pEW of the metal lines. 

An extensive sample of (i)PTF SNe~II was investigated by R15 with special focus on their early-time light curves. R15 established the explosion epochs for 57 events, whose spectra show the strong Balmer P-Cygni profiles typical of SNe~II. Based on the light curves and the spectra of each SN, we subclassified our SNe into SNe~IIP or IIL. Objects with a decline rate ($s_{2}$ in \citealp{anderson14}) $>$ 1.4 mag/100~d during the plateau phase and a low ratio between the EW of H$\alpha$ in absorption and emission were classified as SNe~IIL. In Table~\ref{tab:spectra} we label the SNe~IIL with an asterisk ``*". Only five SNe~IIL belong to our sample of 39 SNe~II.

Here we use the R15 (i)PTF SN sample to investigate the presence of SNe~II at low $Z$, by measuring their pEW$_{5018}$ during the plateau phase. We also check for the correlation between the $Z$ inferred from the pEW measurements and the values obtained by studying the host-galaxy properties.

This Letter is structured as follows.
In Sect.~\ref{sec:obs} the spectral observations of the (i)PTF SN~II sample are presented along with the host-galaxy data. Section~\ref{sec:analysis} describes the EW measurements and the other host-galaxy $Z$ measurements, along with the main results. Our conclusions are summarized in Sect.~\ref{sec:discussion}.

\section{Observations}
\label{sec:obs}

We collected the optical spectra of the 57 SNe~II presented by R15, as obtained by the (i)PTF collaboration. We looked for \ion{Fe}{ii}~$\lambda$5018 in each of the spectra and identified the line in 39 different SNe. For many SNe this line is detected only in a single spectrum, typically the last spectrum obtained during the plateau phase. Even though \ion{Fe}{ii}~$\lambda$5018 can sometimes be detected before the plateau phase, at those early epochs it is not useful for distinguishing between low and high $Z$ (D14), and that is why only 39 out of 57 SNe were analyzed. For the SNe where the line was detected at multiple epochs during the plateau phase, we selected the spectrum with the highest S/N for further analysis. 

The selected spectra were obtained with many different telescopes and instruments, as summarized in Table~\ref{tab:spectra}. Each spectrum has been reduced in the standard manner, including bias and flat-field corrections, wavelength calibration using the spectrum of a comparison lamp, and flux calibration with the spectrum of a standard star observed on the same night. 

For each spectrum where \ion{Fe}{ii}~$\lambda$5018 was identified, we established the phase, based on the explosion date reported by R15. The phase was corrected for time dilation based on the SN redshift (from R15), even though this correction is minimal for our relatively nearby objects. (The average redshift of our sample is $\bar{z}$ $=$ 0.030.) The phase was determined with high accuracy ($\pm1.15$~d on average) given the high cadence of (i)PTF.

We also collected photometry ($ugriz$) and optical spectra of the host galaxies of our SNe from the Sloan Digital Sky Survey \citep[SDSS;][]{ahn14}. Of our 39 SNe, 35 are in the SDSS footprint and have a detected host. An SDSS spectrum is available for only 14 of our galaxies. 
Using these data we are able to independently check the $Z$ estimates from pEW$_{5018}$.

\section{Analysis and results}
\label{sec:analysis}

We measured pEW$_{5018}$ with a MATLAB script based on the formulae given by \citet{nordin11} (see their Eqs. 1 and 2). The uncertainty estimates include the error due to the pseudo-continuum selection and that associated with the noise of the spectrum. The boundaries of the continuum were selected manually with the help of a smoothed spectrum on top of the original data to guide the eye. We compared these EW measurements and uncertainty estimates with those obtained with the IRAF splot EW tool and found that the results were consistent.

We plot pEW$_{5018}$ as a function of SN phase in Fig.~\ref{EW}, also showing the models by D14 for different $Z$. We indicate pEW$_{5018}$ measurements from D14, mainly from CSP SNe~II. Also, the pEW$_{5018}$ values at +50~d presented by \citet{anderson15a} are provided. 
Between $\sim$60 and $\sim$90~d, the pEW$_{5018}$ values inferred for the objects in our sample are on average lower than what was previously presented in the literature. The untargeted nature of the (i)PTF survey, along with its spectroscopic follow-up capability, has allowed us to find a dozen SNe~II (black symbols in Fig.~\ref{EW}) whose 
 pEW$_{5018}$ match spectral models having $Z~=~0.1$~$Z_{\odot}$ (black line in Fig.~\ref{EW}). In some cases these SNe have even smaller
pEW$_{5018}$ than what is expected from these models. Only LSQ13fn \citep{polshaw15} has a comparably small pEW$_{5018}$ (see the magenta empty circle in Fig.~\ref{EW}).
In Fig.~\ref{speclowmet} we show a few examples of (i)PTF SN~II spectra selected among those with small pEW$_{5018}$. This line is particularly faint, but clearly detected given the high S/N of these spectra.

\begin{figure}
\includegraphics[width=9cm]{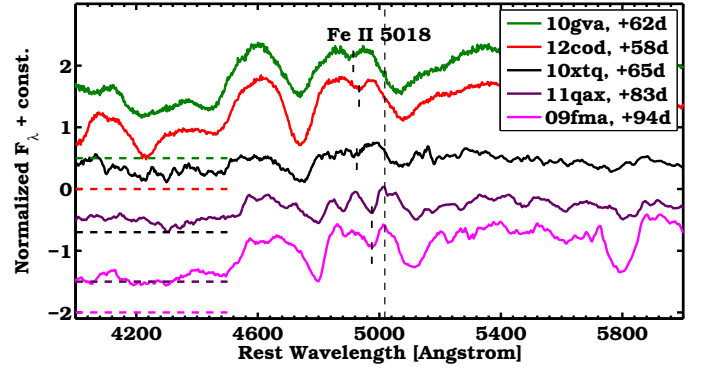}
\label{speclowmet}
\caption{\label{speclowmet}Examples of SN~II spectra with small pEW$_{5018}$. The \ion{Fe}{ii}~$\lambda$5018 rest wavelength is marked by a vertical dashed line, the absorption minima by vertical dashed segments. The spectra are normalized by their median, offset by a constant (dashed horizontal lines), and shown in the rest frame.}
\end{figure}

\setcounter{figure}{3}

\begin{figure}
$\begin{array}{cc}
\includegraphics[width=4.5cm]{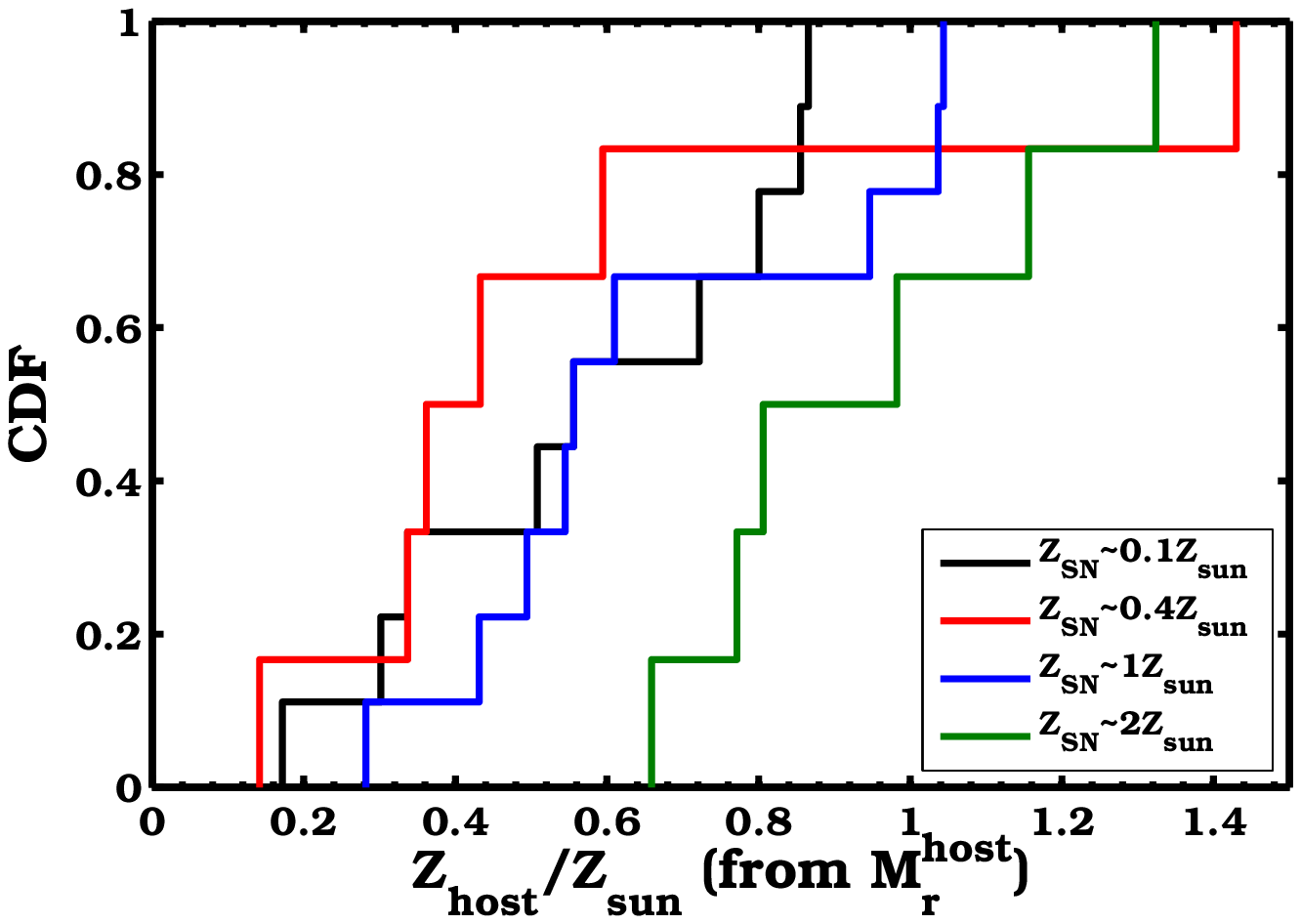}&
\includegraphics[width=4.5cm]{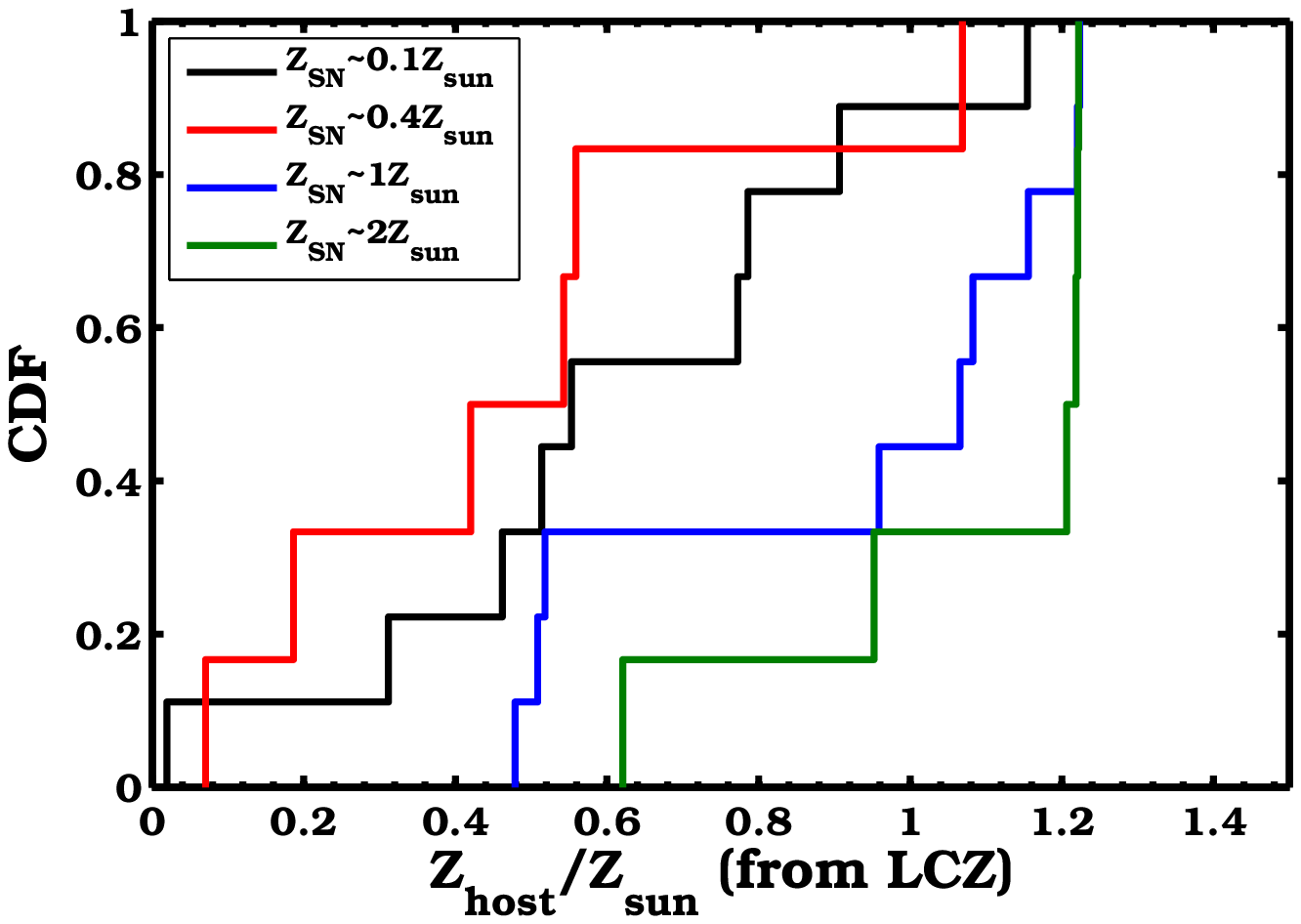}\\
\includegraphics[width=4.5cm]{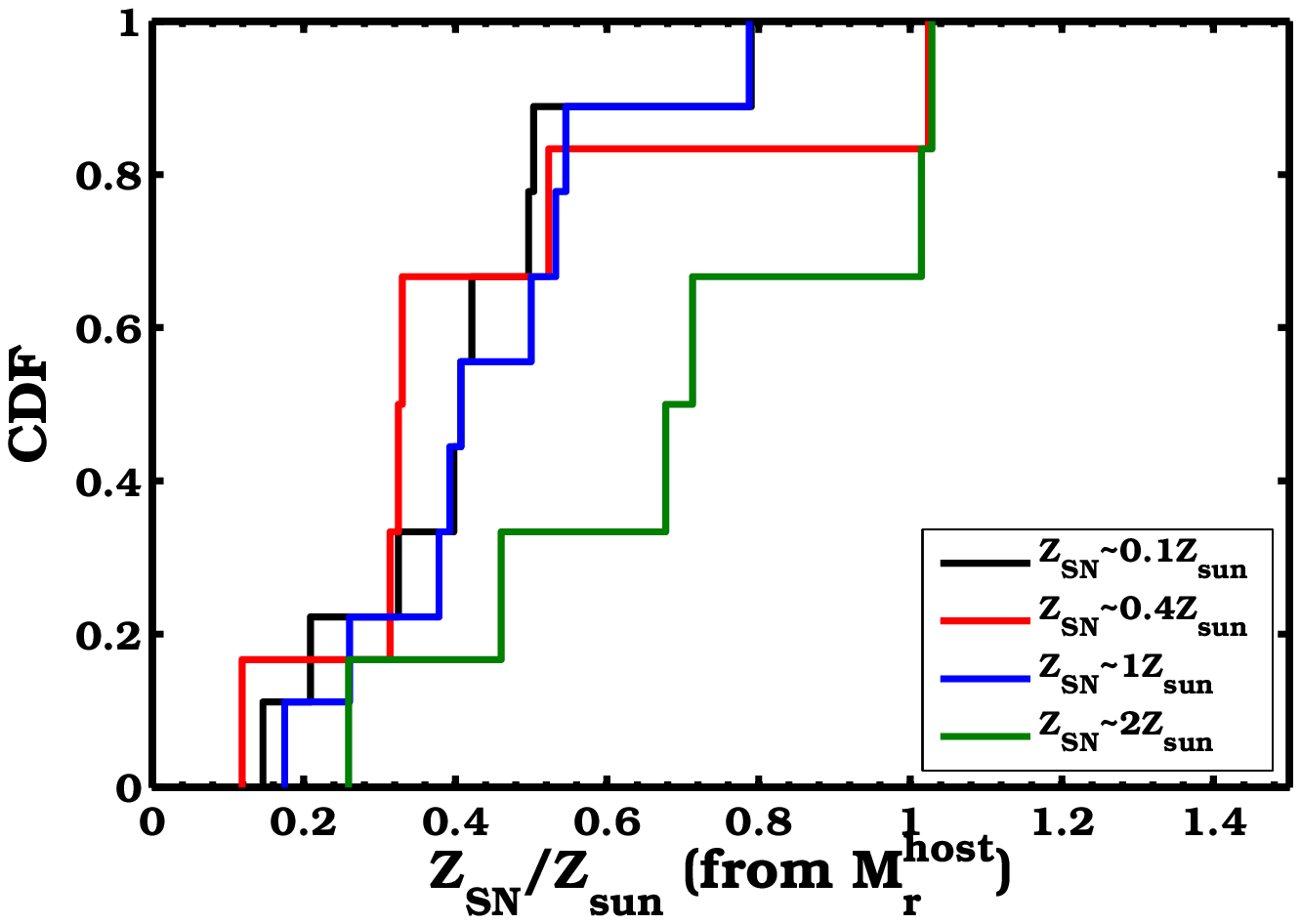}&
\includegraphics[width=4.5cm]{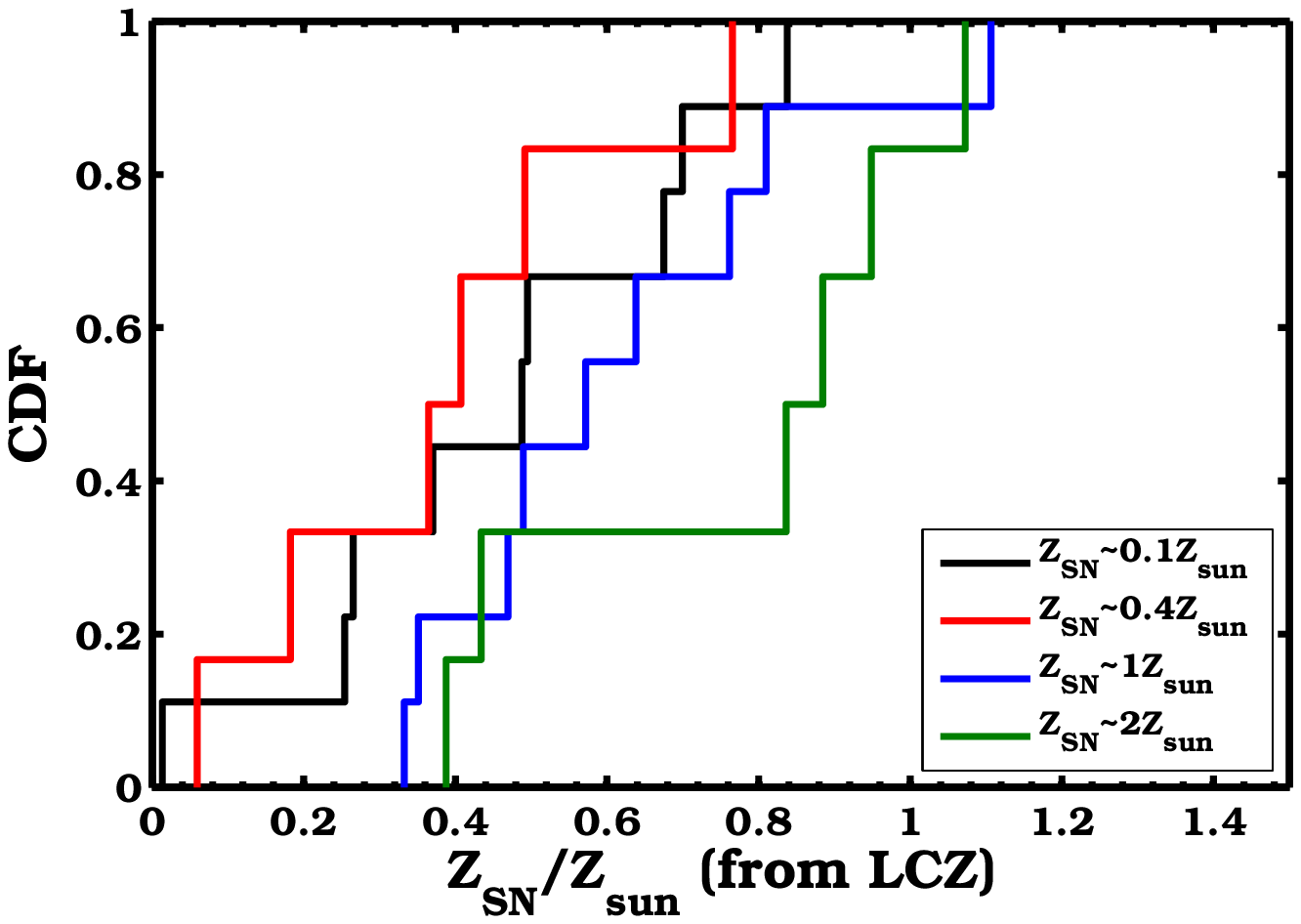}\\
\end{array}$
\caption{ \label{histo1} Top panels: Global metallicity ($Z^{\rm host}$) cumulative distributions from the LZ (left) and the LCZ (right) relations. We subdivided the SNe~IIP into 4 $Z$ bins based on their pEW$_{5018}$ as compared to the models by D14. Bottom panels: as in the top panels, but $Z$ is that at the SN locations, as derived by assuming a (single) $Z$ gradient for all the hosts.} 
\end{figure}

Because of the small number of available host-galaxy spectra, we resorted to using the photometric measurements of the SN host galaxies from SDSS to test whether the SNe with small pEW$_{5018}$ are indeed in small metal-poor galaxies, and if those with large pEW$_{5018}$ are in large, luminous, metal-rich galaxies. First, we converted 
the $r$-band apparent magnitudes from SDSS (Cmodel) to absolute magnitudes ($M_{\rm host}(r)$) using the distance moduli presented by R15 and $E(B-V)_{\rm MW}$ from \citet{ext11}. 
Figure~\ref{pew_vs_mrhost} shows pEW$_{5018}$ versus $M_{\rm host}(r)$ (excluding SNe~IIL). Even if the phases of the spectra span at least two months, there is a correlation between the two observables (Spearman test gives p-value $=$ 0.007). In our sample, SNe with pEW$_{5018}\lesssim$~$-$20~\AA\ never occur in galaxies fainter than $M_{\rm host}(r)~\approx-$19 mag.  
Then, using $M_{\rm host}(r)$, we obtained an estimate of the metal content ($Z_{\rm host}$) for each host via Eq.~1 of \citet{arcavi10}. We plot in Fig.~\ref{histo1} (top-left panel) the cumulative distributions of $Z_{\rm host}$ for the host galaxies of the SNe with pEW$_{5018}$ consistent with $Z_{\rm SN}~\approx~0.1$, 0.4, 1, and 2~$Z_{\odot}$. 

We subdivided our SNe into these four $Z$ bins based on the distance of their pEW$_{5018}$ values from those of the models by D14 (see Fig.~\ref{EW}).
 It indeed seems that SNe with the largest pEW$_{5018}$ at a given phase are in galaxies with the highest amounts of metals (see green line). 
  Since the luminosity-metallicity (LZ) relation from \citet[][see also \citealp{tremonti04}]{arcavi10} is known to be affected by large dispersion,
 we also estimate the host $Z$ via the luminosity-color-metallicity (LCZ) relation by \citet{sanders13}. Making use of their O3N2 calibration along with $M_{\rm host}(g)$ and $(g-r)_{\rm host}$ for each host in order to get the oxygen abundances, these abundances were then converted into $Z_{\rm host}$. In the top-right panel of Fig.~\ref{histo1}, we show that with this improved calibration the SNe with smaller pEW$_{5018}$ (black and red lines) are also located in metal-poorer galaxies. 

To estimate the $Z$ at the location of our SNe within their hosts, we can correct the global $Z$ of their hosts for the metallicity gradient that is known to characterize galaxies \citep[e.g.,][]{p04,taddia13met,taddia15met}, where the nucleus is typically more metal-rich than the outer parts.We used the derived 
global $Z$ as a proxy of the central $Z$, and then adopted an average $Z$ gradient of $-0.47~R_{25}^{-1}$ (see \citealp{p04}). 
The deprojected and radius-normalized distance for each SN from its host center (see Table~\ref{tab:metal}) was estimated using the SN and the host-galaxy coordinates, 
the host-galaxy radius, the host-galaxy axis ratio, and the position angle as obtained from SDSS \citep{ahn14}. 
In Fig.~\ref{histo1} we show the obtained SN location cumulative $Z$ distributions for the four SN groups based on pEW$_{5018}$, using the LZ relation (bottom-left panel) and the LCZ relation (bottom-right panel). With the LCZ calibration, pEW$_{5018}$ is confirmed as a proxy for the actual SN host $Z$, with the objects having smaller pEW$_{5018}$  located at lower $Z$. In all the distributions of Fig.~\ref{histo1}, we only included SNe~IIP, but including SNe~IIL does not change the results significantly.  Spearman tests between $Z_{SN}$ from pEW$_{5018}$ and  $Z_{host}$ (from both LC and LCZ) reveal that there is a correlation with p-values~$<$~0.05. All the $Z$ estimates are reported in Table~\ref{tab:metal}. 

The values of $Z$ from pEW$_{5018}$ cover a wider range than those from LZ and LCZ (See Fig.~\ref{histo1}, and also \citealp{anderson16}, their fig. 10). The latter are obtained from O abundances, whereas pEW$_{5018}$ is essentially a measure of the Fe abundance. The D14 models assume a constant [O/Fe], but at least in the MW
[O/Fe] is lower at higher [Fe/H] (e.g., \citealp{amarsi15}). Therefore, SNe with low pEW$_{5018}$ will be found at higher host-galaxy $Z$ (based on O abundance) than expected from the models based on Fe abundance, and viceversa. 


In Table~\ref{tab:metal}, we also report the $Z$ measurements from the emission lines of the few SDSS galaxy spectra that are available and whose line ratios were consistent with no AGN contamination \citep{baldwin81}.

\begin{figure}
\includegraphics[width=9cm]{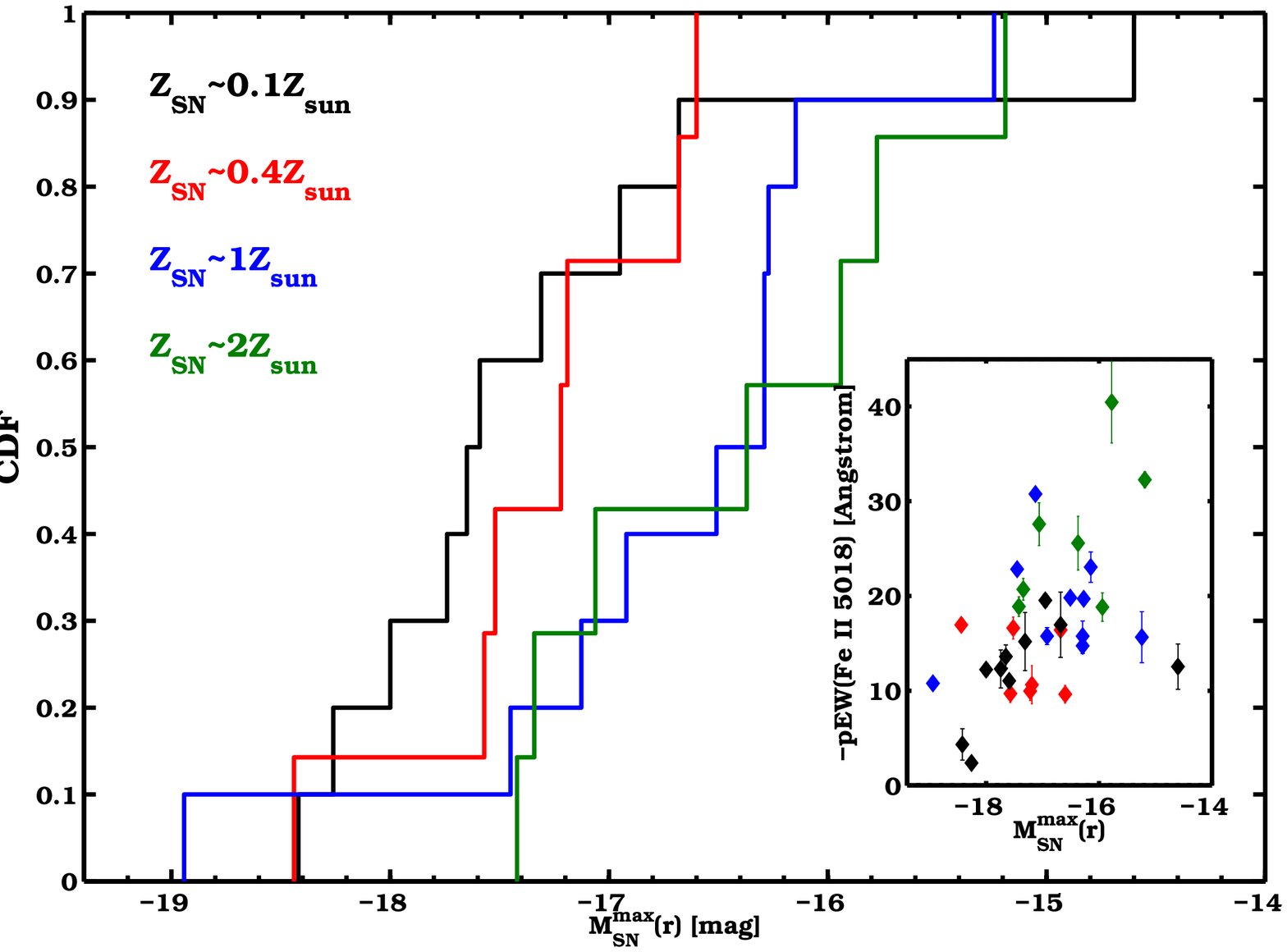}
\caption{ \label{histoMpeak} Cumulative distributions of the SN peak $r$-band absolute magnitude for the four $Z$ bins based on pEW$_{5018}$. SNe~IIP at lower $Z$ tend to be more luminous.}
\end{figure}

We tested if the different SN groups based on pEW$_{5018}$ have different SN observables. The K-S tests show that there is no statistically significant difference among the four groups when we compare the distributions of $r$-band rise time and $r$-band $\Delta m_{15}$. (SN properties were taken from R15.) 
However, we found that 
there is a statistically significant difference between the low- and high-$Z$ SN groups when we compare their absolute $r$-band peak magnitudes [$M_{\rm SN}^{\rm max}(r)$]. These were corrected for the host extinction by measuring the EW of the narrow \ion{Na}{i}~D \citep{turatto03}. SNe at lower $Z$ ($Z~\approx~0.1$; 0.4~$Z_{\odot}$) tend to be more luminous than those at high $Z$ ($Z~\approx~1$; 2~$Z_{\odot}$), with only a 1\% chance of being drawn from the same distribution.
 The $M_{\rm SN}^{\rm max}(r)$ distributions are shown in Fig.~\ref{histoMpeak}. 
The average peak magnitudes of low- and high-$Z$ SNe are  $<M_{\rm SN}^{\rm max}>=-$17.3 mag and $-$16.6 mag, respectively. In the inset of Fig.~\ref{histoMpeak}, we also show that pEW$_{5018}$ measured at different phases during the plateau correlates with the SN peak magnitude. Models of SN~II progenitors with initial mass $=$ 15~$M_{\odot}$ and different $Z$ by \citet{dessart13} show that the $V-$band peak should be fainter for low-$Z$ SNe because they explode with more compact radii, in contrast to our trend. However, their $M_{\rm SN}^{\rm max}(r)$ range is narrower than 1 mag, whereas our observed SNe span 4 mag. 

\section{Conclusions}
\label{sec:discussion}

SNe~IIP were known to occur at relatively high $Z$ (\citealp{anderson10}, D14). Thanks to the untargeted (i)PTF survey, we have shown that SNe~IIP also arise in relatively large numbers from progenitors consistent with  $Z~\approx~0.1$~$Z_{\odot}$. The high quality of the (i)PTF spectra allows us to also measure the weakest \ion{Fe}{ii}~$\lambda$5018 lines. 
The expected trend in pEW($t$) with $Z_{\rm host}$ is observed, although with weak significance. SNe~IIP with smaller pEW tend to occur in metal-poorer environments. Spectral $Z$ measurements are required to better calibrate the relation and assess its dispersion \citep[see, e.g.,][]{anderson15a}. SN~IIP peak magnitudes correlate with $Z$, with more-luminous SNe occurring at lower $Z$. 

\begin{acknowledgements}

We gratefully acknowledge the support from the Knut and Alice Wallenberg Foundation. The Oskar Klein Centre is funded by the Swedish Research Council. A.G.-Y. is supported by the EU/FP7 via ERC grant No. 307260, the Quantum Universe I-Core program by the Israeli Committee for Planning and Budgeting and the ISF; by Minerva and ISF grants; by the Weizmann-UK ``making connections'' program; and by Kimmel and ARCHES awards. A.V.F.'s research is supported by the Christopher R. Redlich Fund, the TABASGO Foundation, and NSF grant AST-1211916. We are grateful to the staffs at the many observatories where data for this study were collected (Palomar, Lick, Keck, etc.). We thank R. Foley, J. Bloom, Green, N. Cucchiara, A. Horesh, K. I. Clubb, M. T. Kandrashoff, K. Maguire, A. De Cia, S. Tang, B. Zackay, B. Sesar, A. Waszczak, I. Shivvers, who helped with some of the observations and data reduction. Research at Lick Observatory is partially supported by a generous gift
from Google. Some of the data presented herein were obtained at the
W. M. Keck Observatory, which is operated as a scientific partnership
among the California Institute of Technology, the University of
California, and NASA; the observatory was made possible by the
generous financial support of the W. M. Keck Foundation. This research used resources of the National Energy Research Scientific Computing Center, a DOE Office of Science User Facility supported by the Office of Science of the U.S. Department of Energy under Contract No. DE-AC02-05CH11231. J.M.S. is supported by an NSF Astronomy and Astrophysics Postdoctoral Fellowship under award AST-1302771. D.X. acknowledges the support of the One-Hundred-Talent Program from the National Astronomical Observatories, Chinese Academy of Sciences. This work is partly based on the Bachelor thesis by P. Moquist.

\end{acknowledgements}

\onecolumn

\setcounter{figure}{2}
\begin{figure}
\centering
\includegraphics[width=9cm]{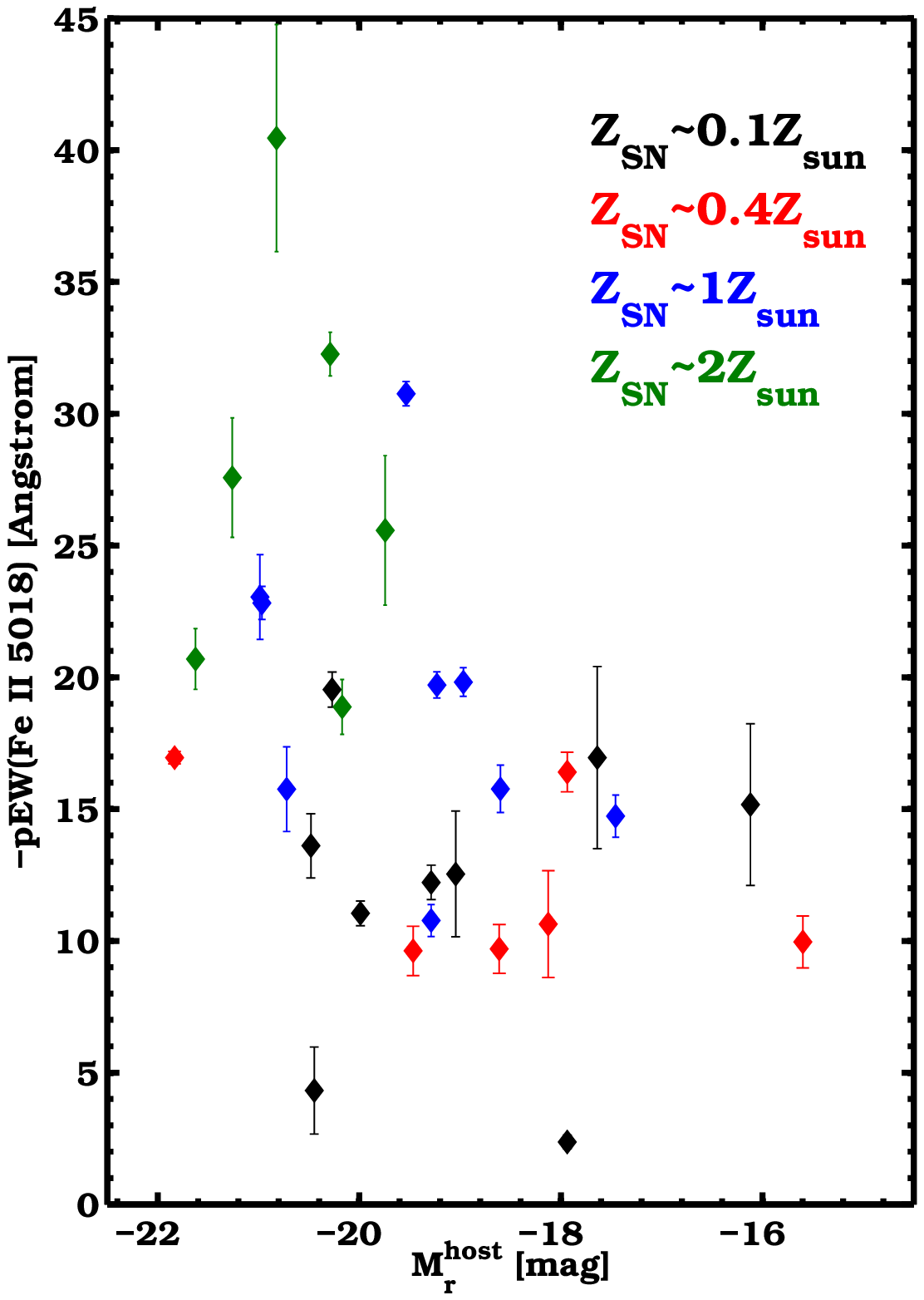}
\caption{ \label{pew_vs_mrhost}pEW$_{5018}$ versus $M_{r}^{host}$. Symbols are color-coded as in Fig.~\ref{EW}.} 
\end{figure}

\begin{deluxetable}{lllcl}
\tabletypesize{\scriptsize}
\tablewidth{0pt}
\tablecaption{Log of spectral observations and pEW measurements.\label{tab:spectra}}
\tablehead{
\colhead{(i)PTF SN}&
\colhead{Phase\tablenotemark{a}}&
\colhead{$-$pEW$_{5018}$}&
\colhead{Telescope+Instrument}&
\colhead{Date}\\
\colhead{}&
\colhead{(d)}&
\colhead{(\AA)}&
\colhead{}&
\colhead{(UT)}}
\startdata
09ecm     &  41.87(0.35) & 18.88(1.04)   &      Keck1+LRIS     & 2009 Oct 23 \\        
09fma     &  93.66(1.05) & 12.31(2.01)   &      P200+DBSP      & 2010 Jan 09 \\        
10bgl     &  34.35(0.97) & 20.70(1.15)   &      Keck1+LRIS     & 2010 Feb 06 \\        
10gva     &  62.17(0.88) & 2.37(0.26)    &      Keck1+LRIS     & 2010 Jun 12 \\        
10gxi     &  38.04(2.41) & 10.64(2.02)   &      P200+DBSP      & 2010 Jun 13 \\        
10jwr     &  28.03(2.35) & 10.77(0.60)   &      Keck1+LRIS     & 2010 Jul 07 \\        
10mug*     &  45.46(2.35) & 10.72(1.28)   &      P200+DBSP      & 2010 Aug 14 \\        
10osr     &  88.53(0.91) & 15.17(3.07)   &      Lick 3-m+Kast  & 2010 Oct 11 \\        
10pjg     &  56.91(2.39) & 15.75(1.60)   &      P200+DBSP      & 2010 Sep 06 \\        
10qwz     &  62.87(1.47) & 23.04(1.61)   &      Lick 3-m+Kast  & 2010 Oct 11 \\        
10rem     &  62.89(1.87) & 16.41(0.75)   &      Keck2+DEIMOS   & 2010 Oct 12 \\        
10vdl     &  53.84(1.95) & 15.65(2.69)   &      Keck2+DEIMOS   & 2010 Nov 07 \\        
10xtq     &  64.82(0.45) & 4.32(1.65)    &      P200+DBSP      & 2010 Dec 06 \\        
11ajz     &  99.12(0.94) & 13.61(1.22)   &      Lick 3-m+Kast  & 2011 May 13 \\        
11go      &  57.55(1.37) & 25.57(2.84)   &      P200+DBSP      & 2011 Mar 10 \\        
11hsj     &  77.60(1.90) & 16.61(1.16)   &      Lick 3-m+Kast  & 2011 Sep 29 \\        
11htj     &  110.21(1.46) & 16.96(3.45)  &      P200+DBSP      & 2011 Oct 30 \\        
11iqb     &  64.97(0.21) & 16.95(0.24)   &      Keck1+LRIS     & 2011 Sep 26 \\        
11izt     &  37.34(2.41) & 18.82(1.51)   &      WHT+ISIS       & 2011 Aug 31 \\        
11qax     &  83.47(0.40) & 12.22(0.65)   &      KPNO4m+RC Spec & 2012 Jan 26 \\        
12bro     &  44.18(0.37) & 9.96(0.99)    &      P200+DBSP      & 2012 Apr 29 \\        
12bvh     &  89.80(0.95) & 30.76(0.46)   &      Lick 3-m+Kast  & 2012 Jun 14 \\        
12cod*     &  57.91(1.88) & 3.76(0.22)    &      TNG+DOLORES    & 2012 May 31 \\        
12fip     &  38.47(0.94) & 9.62(0.93)    &      P200+DBSP      & 2012 Jul 21 \\        
12fo      &  116.10(0.96) & 19.53(0.66)  &      Keck1+LRIS     & 2012 Apr 29 \\        
12ftc     &  41.60(0.90) & 9.69(0.93)    &      P200+DBSP      & 2012 Jul 27 \\        
12gnn*     &  42.30(0.94) & 19.29(1.59)   &      WHT+ISIS       & 2012 Aug 21 \\        
12hsx     &  46.21(0.25) & 15.77(0.90)   &      WHT+ISIS       & 2012 Aug 21 \\        
13aaz     &  74.24(1.42) & 32.27(0.83)   &      P200+DBSP      & 2013 Jun 03 \\        
13akg     &  58.19(2.42) & 19.82(0.54)   &      Keck2+DEIMOS   & 2013 Jun 06 \\        
13bjx     &  62.72(0.47) & 22.82(0.62)   &      P200+DBSP      & 2013 Aug 03 \\        %
13bld     &  33.95(0.45) & 27.58(2.27)   &      P200+DBSP      & 2013 Jul 05 \\        
13cnk     &  56.82(0.44) & 19.71(0.50)   &      Keck2+DEIMOS   & 2013 Oct 04 \\        
13dkk     &  74.48(0.35) & 12.54(2.39)   &      P200+DBSP      & 2013 Nov 26 \\        
13dkz     &  49.73(0.44) & 14.73(0.80)   &      P200+DBSP      & 2013 Nov 02 \\        
13dla*     &  75.52(0.43) & 16.50(5.79)   &      Keck1+LRIS     & 2013 Dec 02 \\        
13dqy     &  91.12(0.44) & 11.04(0.47)   &      P200+DBSP      & 2014 Jan 06 \\        
14adz*     &  64.94(0.33) & 14.75(0.87)   &      Keck1+LRIS     & 2014 May 28 \\        
14aoi     &  68.05(0.05) & 40.46(4.30)   &      Lick 3-m+Kast  & 2014 Jun 30 \\        

\enddata
\tablenotetext{a}{From explosion date, and corrected for time dilation.}
\tablenotetext{*}{SN~IIL. The other objects are SNe~IIP.}
\end{deluxetable}

\begin{deluxetable}{lllllll|lll|lll}
\rotate
\tabletypesize{\scriptsize}
\tablewidth{0pt}
\tablecaption{Log of metallicity estimates and galaxy properties for our sample of (i)PTF SNe~II\label{tab:metal}}
\tablehead{
\colhead{(i)PTF SN}&
\colhead{Host galaxy}&
\colhead{$\alpha_{\rm host}$ (J2000)}&
\colhead{$\delta_{\rm host}$ (J2000)}&
\colhead{$d_{\rm SN}$/$R_{g}$}&
\colhead{$M_{\rm host}(r)$}&
\colhead{$(g-r)_{\rm host}$}&
\colhead{$Z_{\rm host}$}&
\colhead{$Z_{\rm host}$}&
\colhead{$Z_{\rm host}$}&
\colhead{$Z_{\rm SN}$}&
\colhead{$Z_{\rm SN}$}&
\colhead{$Z_{\rm SN}$}\\
\colhead{}&
\colhead{}&
\colhead{(hh:mm:ss)}&
\colhead{(dd:mm:ss)}&
\colhead{}&
\colhead{(mag)}&
\colhead{(mag)}&
\colhead{($Z_{\odot}$; LZ)}&
\colhead{($Z_{\odot}$; LCZ)}&
\colhead{($Z_{\odot}$; O3N2)}&
\colhead{($Z_{\odot}$; LZ)}&
\colhead{($Z_{\odot}$; LCZ)}&
\colhead{($Z_{\odot}$; O3N2)}}
\startdata
09ecm  &  2MASX J01064313-0622468       &       01:06:42.914   &      -06:22:44.83      &   0.40      &  -20.17  & 0.48        &  0.77  & 0.95    &    \ldots    &    0.68     & 0.84   &   \ldots   \\
09fma  &  GALEXASC J031023.07-100001.5   &       03:10:23.161  &       -10:00:01.04     &   \ldots    &  \ldots  & \ldots      &  \ldots   & \ldots     &    \ldots    &    \ldots      & \ldots    &   \ldots   \\
10bgl  &  NGC 3191                      &       10:19:05.131   &      +46:27:14.77      &   0.78      &  -21.63  & 0.60        &  1.32  & 1.22    &    \ldots    &    1.03     & 0.95   &   \ldots   \\
10gva  &  SDSS J122355.39+103448.9      &       12:23:55.395   &      +10:34:48.97      &   0.11      &  -17.94  & 0.44        &  0.34  & 0.51    &    0.59   &    0.32     & 0.50   &   0.57  \\
10gxi  &  SDSS J124433.49+310505.7      &       12:44:33.497   &      +31:05:05.75      &   0.43      &  -18.13  & 0.36        &  0.36  & 0.42    &    0.50   &    0.31     & 0.36   &   0.43  \\
10jwr  &  SDSS J161215.97+320414.6      &       16:12:15.963   &      +32:04:14.68      &   0.13      &  -19.29  & 0.73        &  0.56  & 1.16    &    \ldots    &    0.53     & 1.11   &   \ldots   \\
10mug*  &  SDSS J150406.83+282917.6             &       15:04:06.831   &      +28:29:17.68      &   0.15      &  -16.96  & 0.42        &  0.24  & 0.35    &    \ldots    &    0.22     & 0.33   &   \ldots   \\
10osr  &  SDSS J234545.05+112842.1      &       23:45:45.055   &      +11:28:42.16      &   0.50      &  -16.12  & 0.48        &  0.17  & 0.31    &    \ldots    &    0.15     & 0.27   &   \ldots   \\
10pjg  &  Zw 104 NED02                  &       23:23:08.856   &      +13:02:44.27      &   1.97      &  -20.72  & 0.51        &  0.95  & 1.08    &    \ldots    &    0.50     & 0.57   &   \ldots   \\
10qwz  &  UGC 12687                     &       23:35:17.579   &      +12:55:26.42      &   2.00      &  -20.99  & 0.76        &  1.04  & 1.22    &    \ldots    &    0.55     & 0.64   &   \ldots   \\
10rem  &  SDSS J171743.62+205230.4      &       17:17:43.627   &      +20:52:30.43      &   0.07      &  -17.94  & 0.17        &  0.34  & 0.19    &    \ldots    &    0.33     & 0.18   &   \ldots   \\
10vdl  &  -                              &       \ldots        &     \ldots             &   \ldots    &  \ldots  & \ldots      &  \ldots   & \ldots     &    \ldots    &    \ldots      & \ldots    &   \ldots   \\
10xtq  &  SDSS J082314.14+215755.4      &       08:23:14.148   &      +21:57:55.44      &   0.24      &  -20.45  & 0.42        &  0.86  & 0.91    &    \ldots    &    0.79     & 0.84   &   \ldots   \\
11ajz  &  IC 2373                       &       08:26:48.979   &      +20:21:53.42      &   2.40      &  -20.48  & 0.20        &  0.87  & 0.55    &    0.77   &    0.40     & 0.25   &   0.36  \\
11go   &  SDSS J113200.20+534250.5      &       11:32:00.210   &      +53:42:50.57      &   1.10      &  -19.74  & 0.32        &  0.66  & 0.62    &    0.68   &    0.46     & 0.43   &   0.48  \\
11hsj  &  GALEXASC J165757.67+551105.0   &       16:57:57.673  &       +55:11:05.04      &  \ldots    &   \ldots  & \ldots     &  \ldots   & \ldots     &    \ldots    &    \ldots      & \ldots    &   \ldots   \\
11htj  &  SDSS J211603.12+123124.4      &       21:16:03.128   &      +12:31:24.41      &   1.13      &  -17.64  & -0.20       &  0.30  & 0.02    &    \ldots    &    0.21     & 0.01   &   \ldots   \\
11iqb  &  NGC 0151                      &       00:34:02.791   &      -09:42:19.02      &   1.03      &  -21.83  & 0.87        &  1.43  & 1.07    &    \ldots    &    1.02     & 0.77   &   \ldots   \\
11izt  &  GALEXASC J015226.07+353023.6   &       01:52:26.074  &       +35:30:23.66      &  \ldots    &   \ldots  & \ldots     &  \ldots   & \ldots     &    \ldots    &    \ldots      & \ldots    &   \ldots   \\
11qax  &  SDSS J234226.03+001521.5      &       23:42:26.030   &      +00:15:21.57      &   0.31      &  -19.29  & 0.46        &  0.56  & 0.77    &    0.75   &    0.50     & 0.70   &   0.68  \\
12bro  &  SDSS J122417.03+185529.5      &       12:24:17.037   &      +18:55:29.50      &   0.55      &  -15.60  & 0.22        &  0.14  & 0.07    &    \ldots    &    0.12     & 0.06   &   \ldots   \\
12bvh  &  M95                           &       10:43:57.691   &      +11:42:13.70      &   3.84      &  -19.54  & 0.83        &  0.61  & 1.22    &    \ldots    &    0.18     & 0.35   &   \ldots   \\
12cod*  &  SDSS J132232.47+544905.5             &       13:22:32.477   &      +54:49:05.53      &   1.67      &  -19.47  & 0.17        &  0.60  & 0.35    &    0.85   &    0.35     & 0.21   &   0.50  \\
12fip  &  SDSS J150050.86+092027.6      &       15:00:50.863   &      +09:20:27.64      &   0.39      &  -19.47  & 0.31        &  0.59  & 0.56    &    \ldots    &    0.52     & 0.49   &   \ldots   \\
12fo   &  SDSS J125837.28+271035.8      &       12:58:37.282   &      +27:10:35.81      &   1.47      &  -20.27  & 0.36        &  0.80  & 0.79    &    1.03   &    0.50     & 0.49   &   0.64  \\
12ftc  &  SDSS J150501.72+200554.6      &       15:05:01.730   &      +20:05:54.68      &   0.89      &  -18.61  & 0.39        &  0.43  & 0.54    &    \ldots    &    0.32     & 0.41   &   \ldots   \\
12gnn*  &  SDSS J155849.24+361010.4             &       15:58:49.242   &      +36:10:10.42      &   0.16      &  -15.82  & 0.45        &  0.15  & 0.24    &    \ldots    &    0.15     & 0.23   &   \ldots   \\
12hsx  &  SDSS J005503.33+421954.0      &       00:55:03.337   &      +42:19:54.06      &   0.18      &  -18.60  & 0.37        &  0.43  & 0.52    &    \ldots    &    0.41     & 0.49   &   \ldots   \\
13aaz  &  M65                           &       11:18:55.910   &      +13:05:32.33      &   3.50      &  -20.29  & 0.87        &  0.81  & 1.21    &    \ldots    &    0.26     & 0.39   &   \ldots   \\
13akg  &  SDSS J113437.02+545328.7      &       11:34:37.021   &      +54:53:28.77      &   0.71      &  -18.97  & 0.61        &  0.49  & 0.96    &    \ldots    &    0.39     & 0.76   &   \ldots   \\
13bjx  &  SDSS J141451.26+364723.8      &       14:14:51.260   &      +36:47:23.83      &   0.85      &  -20.97  & 0.47        &  1.04  & 1.07    &    \ldots    &    0.79     & 0.81   &   \ldots   \\
13bld  &  SDSS J162454.16+410302.8      &       16:24:54.163   &      +41:03:02.74      &   0.40      &  -21.26  & 0.64        &  1.16  & 1.22    &    \ldots    &    1.01     & 1.07   &   \ldots   \\
13cnk  &  SDSS J020213.41+075831.3      &       02:02:13.413   &      +07:58:31.27      &   1.12      &  -19.23  & 0.28        &  0.54  & 0.48    &    \ldots    &    0.38     & 0.33   &   \ldots   \\
13dkk  &  NGC 7732                      &       23:41:33.804   &      +03:43:29.53      &   0.68      &  -19.04  & 0.29        &  0.51  & 0.46    &    \ldots    &    0.41     & 0.37   &   \ldots   \\
13dkz  &  SDSS J013611.64+333703.6      &       01:36:11.645   &      +33:37:03.65      &   0.25      &  -17.46  & 0.49        &  0.28  & 0.51    &    \ldots    &    0.26     & 0.47   &   \ldots   \\
13dla*  &  SDSS J010249.10-004430.2             &       01:02:49.101   &      -00:44:30.25      &   0.42      &  -17.02  & 0.32        &  0.24  & 0.24    &    \ldots    &    0.21     & 0.21   &   \ldots   \\
13dqy  &  NGC 7610                      &       23:19:41.376   &      +10:11:06.04      &   1.66      &  -19.99  & 0.66        &  0.72  & 1.15    &    \ldots    &    0.42     & 0.67   &   \ldots   \\
14adz*  &  SDSS J134957.67+374508.2             &       13:49:57.671   &      +37:45:08.27      &   0.22      &  -19.50  & 0.34        &  0.60  & 0.61    &    \ldots    &    0.56     & 0.57   &   \ldots   \\
14aoi  &  NGC 4134                      &       12:09:10.010   &      +29:10:36.86      &   0.99      &  -20.82  & 0.68        &  0.98  & 1.22    &    \ldots    &    0.71     & 0.88   &   \ldots   \\                     
\enddata
\tablenotetext{*}{SN~IIL. The other objects are SNe~IIP.}
\tablecomments{For each SN we report the host-galaxy name, the host-galaxy coordinates, the deprojected distance of each SN from its host-galaxy center (normalized by the De Vaucouleurs $g$-band radius of the host), and the absolute $r$-band magnitude as well as $g-r$ color of the host galaxy. In the last six columns $Z$ is reported for the host and at the SN position as measured from three different methods: (1) the luminosity-metallicity relation (LZ), (2) the luminosity-color-metallicity relation (LCZ), and (3) the strong-line diagnostic (O3N2; \citealp{pettini04}).
When computing the metallicity at the SN distance from the host-galaxy center, we converted the average $Z$ gradient in units of DeVRad$\_$g$^{-1}$ by multiplying the metallicity gradient by 0.3, which is the typical ratio between DeVRad$\_$g$^{-1}$ and the isophotal radius in the $g$ band. The isophotal radius, isoA$\_$g, is available only for the SDSS galaxies in the DR7. 
The host of PTF10vdl is not detected, those of PTF09fma, PTF11hsj, and PTF11hzt are not in the SDSS footprint. We exclude these hosts from the analysis presented in Fig.~\ref{histo1}. The vast majority of our SN hosts are in the optimal $M_g-\alpha(g-r)$
 range where the LCZ relation was calibrated. The adoption of a standard gradient for all the host galaxies 
 is a strong assumption and could lead to erroneous local metallicity estimates for some of the SNe if their
  actual gradient is significantly different from ours, as it could be for dwarf galaxies.}
\end{deluxetable}



\begin{thebibliography}{100}

\expandafter\ifx\csname natexlab\endcsname\relax\def\natexlab#1{#1}\fi


\bibitem[Ahn et al.(2014)]{ahn14} Ahn, C.~P., Alexandroff, 
R., Allende Prieto, C., et al.\ 2014, \apjs, 211, 17 

\bibitem[Amarsi et al.(2015)]{amarsi15} Amarsi, A.~M., Asplund, 
M., Collet, R., \& Leenaarts, J.\ 2015, \mnras, 454, L11 

\bibitem[Anderson et al.(2016)]{anderson16} Anderson, J.~P., 
Gutierrez, C.~P., Dessart, L., et al.\ 2016, arXiv:1602.00011 

\bibitem[Anderson et al.(2010)]{anderson10} Anderson, J.~P., 
Covarrubias, R.~A., James, P.~A., Hamuy, M., 
\& Habergham, S.~M.\ 2010, \mnras, 407, 2660 

\bibitem[Anderson et al.(2014)]{anderson14} Anderson, J.~P., 
Gonz{\'a}lez-Gait{\'a}n, S., Hamuy, M., et al.\ 2014, \apj, 786, 67 

\bibitem[Anderson et al.(2015a)]{anderson15a} Anderson, J.~P., Guti{\'e}rrez, C.~P., \& Dessart, L.\ 2015, arXiv:1510.04271 

\bibitem[Arcavi et al.(2010)]{arcavi10} Arcavi, I., Gal-Yam, A., 
Kasliwal, M.~M., et al.\ 2010, \apj, 721, 777 

\bibitem[Baldwin et al.(1981)]{baldwin81} Baldwin, J.~A., 
Phillips, M.~M., \& Terlevich, R.\ 1981, \pasp, 93, 5 (BPT)

\bibitem[Dessart et al.(2014)]{dessart14} Dessart, L., Gutierrez, 
C.~P., Hamuy, M., et al.\ 2014, \mnras, 440, 1856 [D14]

\bibitem[Dessart et al.(2013)]{dessart13} Dessart, L., Hillier, 
D.~J., Waldman, R., \& Livne, E.\ 2013, \mnras, 433, 1745 

\bibitem[Filippenko et al.(1997)]{filippenko97} Filippenko, A. V.
  1997, ARAA, 35, 309

\bibitem[Gal-Yam et al.(2011)]{galyam11} Gal-Yam, A., Kasliwal, 
M.~M., Arcavi, I., et al.\ 2011, \apj, 736, 159 

\bibitem[Li et al.(2011)]{li11} Li, W., Chornock, R., 
Leaman, J., et al.\ 2011, \mnras, 412, 1473 

\bibitem[Nordin et 
al.(2011)]{nordin11} Nordin, J., {\"O}stman, L., Goobar, A., et al.\ 2011, \aap, 526, A119 

\bibitem[Pettini 
\& Pagel(2004)]{pettini04} Pettini, M., \& Pagel, B.~E.~J.\ 2004, \mnras, 348, L59 

\bibitem[Pilyugin et 
al.(2004)]{p04} Pilyugin, L.~S., V{\'{\i}}lchez, J.~M., \& Contini, T.\ 2004, \aap, 425, 849


 \bibitem[Polshaw et al.(2015)]{polshaw15} Polshaw, J., Kotak, R., 
Dessart, L., et al.\ 2015, arXiv:1511.01718 

\bibitem[Rubin et al.(2015)]{rubin16} Rubin, A., Gal-Yam, A., 
De Cia, A., et al.\ 2015, arXiv:1512.00733 [R15]

\bibitem[Sanders et al.(2013)]{sanders13} Sanders, N.~E., 
Levesque, E.~M., \& Soderberg, A.~M.\ 2013, \apj, 775, 125 

\bibitem[Schlafly 
\& Finkbeiner(2011)]{ext11} Schlafly, E.~F., \& Finkbeiner, D.~P.\ 2011, \apj, 737, 103 

\bibitem[Smartt(2009)]{smartt09} Smartt, S.~J.\ 2009, \araa, 47, 63 

\bibitem[Taddia et al.(2015)]{taddia15met} Taddia, F., Sollerman, J., Fremling, C., et al.\ 2015, \aap, 580, A131  

\bibitem[Taddia et al.(2013)]{taddia13met} Taddia, F., Sollerman, J., Razza, A., et al.\ 2013, \aap, 558, A143 

\bibitem[Tremonti et al.(2004)]{tremonti04} Tremonti, C.~A., 
Heckman, T.~M., Kauffmann, G., et al.\ 2004, \apj, 613, 898 

\bibitem[Turatto et al.(2003)]{turatto03} Turatto, M., Benetti, 
S., 
\& Cappellaro, E.\ 2003, From Twilight to Highlight: The Physics of Supernovae, 200 

\end{thebibliography}
\end{document}